\documentclass[a4paper,11pt]{article}
\usepackage{jheppub} % for details on the use of the package, please
\usepackage[T1]{fontenc} % if needed
\title{The Entropy of BTZ Black Hole from Loop Quantum Gravity}

\author[a]{Jingbo Wang}

\affiliation[a]{Department of Physics, Beijing Normal University, \\ Beijing, 100875, People's Republic of China}

\emailAdd{shuijing@mail.bnu.edu.cn}

\abstract{In this paper, we calculated the entropy of the BTZ black hole in the framework of loop quantum gravity. We got the result that the horizon degrees of freedom can be described by the 2D SO(1,1) punctured BF theory. Finally we got the area law for the entropy of BTZ black hole.}

\begin{document}

\maketitle
\flushbottom

\section{Introduction}
Black hole has attract people's attention since a long time ago. The pioneering work of Bekenstein\cite{bk1}, Hawking\cite{hawk1} and others \cite{bch1}during the seventies of last century have suggested that black hole have temperature and entropy. The entropy is given by the famous Bekenstein-Hawking area law. Understanding those properties is a fundamental challenge of quantum gravity.

Loop quantum gravity\cite{rov1,thie1,ash4,ma1} is a proposal for the theory of quantum gravity. The black hole entropy calculation in LQG is based on the effective description of quantum gravitational degree of freedom at the black horizon obtained from quantization of the classical phase space describing the isolated horizon\cite{ash1}. In those models, the degree of freedom at the horizon is described by the Chern-Simons theory with the SU(2)\cite{enpp1} (or U(1)\cite{ash1}) gauge group.

BTZ black hole\cite{btz1} is a solution of Einstein field equation with negative cosmological constant in 3 dimension space-time. Since it is a black hole, it has temperature and entropy as follows:
\begin{equation}\label{1}
    T=\frac{\kappa}{2\pi},\, S=\frac{2 \pi r_+}{4G}.
\end{equation}
where $\kappa$ is the surface gravity, and $r_+$ is the radius of the outer horizon. There are many explanation for this entropy, such as AdS/CFT correspondence\cite{ads1}, Chern-Simons theory\cite{ads2}, spin foam model\cite{spin2,spin1} and so on. For a review, see\cite{ads2}. But up to now, there are fewer  explanation in the framework of LQG.

Just recently, E. Frodden et al\cite{frod1} calculated the entropy of the black hole in the context of 3D Euclidean loop quantum gravity. They character the black hole as a $U_q(su(2))$ quantum spin network state, and calculate the number of SU(2) intertwiners between the representation $j_l$ to give the entropy of the black hole.

Our work is along the traditional ABCK method\cite{ash1}. The symplectic form of black hole can be split into bulk and boundary term, and we conclude that the boundary degrees of freedom can be described by an effective topological field theory. The dimension of the boundary Hilbert space give the entropy of the black hole.

The article is organized as follows. In section 2 we will give the symplectic structure which split into two terms. In section 3 we will show that the 2D BF theory can give the same symplectic form as the boundary term, so we concluded that the boundary degrees of freedom in black hole can be described by a 2D BF theory on the isolated horizon. In section 4, we summary some results for the loop quantum gravity in 2+1 dimension and count the number of the permissible states to show that the entropy of BTZ black hole obey the area law. We will close with a discussion of our results in section 5. In this paper, we choose $\hbar=c=1$.
\section{The Symplectic Structure}
Since the standard definition of black hole is a global definition, a more better concept is the isolated horizon\cite{ash3}. The isolated horizon contain the usual black hole and the cosmological horizons.

Let $\mathcal{M}$ be a three dimensional manifold with metric tensor $g_{ab}$ of signature $(-, +, +)$. An isolated horizon is a null hypersurface $\Delta \subset \mathcal{M}$ with topology $S\times \mathbb{R}$. The pull-back of the gravitational fields to $\Delta$ satisfying the isolated horizon boundary conditions\cite{ash3}.

We will use the first order framework based on orthonormal co-triads $e_I$ and SO(2,1) connections $A^I$ where $I$ takes values in the Lie algebra of SO(2,1). The action for 2+1-dimensional gravity theory is given by\cite{ash2}:
\begin{equation}\label{2}
    S(e,A)=\frac{1}{8\pi G}\int_{\mathcal{M}} (e^I\wedge F_I-\frac{\Lambda}{6} \varepsilon^{IJK} e_I \wedge e_J \wedge e_K)-\frac{1}{16\pi G} \int_{\tau_{\infty}} e^I \wedge A_I,
\end{equation}
where the $F_I$ is the curvature of the gravitational connection $A_I$ and $\tau_{\infty}$ denotes the boundary at infinity. We choose the boundary condition at infinity as same as paper\cite{ash2}, that is, the infinity of space-time is diffeomorphic to the infinity of the BTZ space-time.

The covariant phase space $\Gamma$ will consist of solutions $(e^I, A_I)$ to the Einstein equation, satisfying some boundary condition. As usual, to construct the symplectic structure on $\Gamma$, we begins with the anti-symmetrized second variation of the action (\ref{2}). Applying the equation of motion to the second variation, one find that the integral over $\mathcal{M}$ reduces to surface term at $M^{\pm}$ and at $\Delta$. The surface term at $\tau_{\infty}$ vanishes because of the asymptotic fall-off condition. The relevant term for the boundary symplectic form is given by:
\begin{equation}\label{3}
    \frac{1}{8\pi G}\int_{\Delta} \delta_2 \underline{e^I} \wedge \delta_1 \underline{A_I}
\end{equation}
where the underline denote the pull back to the isolated horizon $\Delta$.

At this stage, we need the special properties of the isolated horizon. We use the Newman-Penrose null tetrad formula which consisting of two null 1-form $l_a$ and $n_a$, and a space-like 1-form $m_a$. An orthonormal co-triad is given by\cite{ash2}
\begin{equation}\label{4}
    e^I_a=-n^I l_a-l^I n_a +m^I m_a.
\end{equation}
The connection 1-form is given by
\begin{equation}\label{5}
    A^I_a=(\pi n_a+\nu l_a-\mu m_a)l^I+(\kappa_{NP}n_a+\tau l_a-\rho m_a)n^I+(-\epsilon n_a-\gamma l_a+\alpha m_a)m^I.
\end{equation}
where the $\alpha, \gamma \cdots$ are the spin-coefficients. On an isolated horizon, the 1-form $l_a$ is zero, and the spin-coefficients are subject to some constrains, such as $\rho\triangleq 0 \triangleq \kappa_{NP}$ and so on, where $\triangleq$ means equal on the isolated horizon. Inserting those properties into the (\ref{4}) and (\ref{5}), we will get the pull-back of the co-triad and the connection 1-form to the isolated horizon as follows:
\begin{equation}\label{6}
    \underline{e^I_a}= -l^I n_a +m^I m_a,\,
    \underline{A^I_a}=(\pi l^I-\epsilon m^I) n_a + (\alpha m^I-\mu l^I)m_a.
\end{equation}

Let's return to the equation (\ref{4}). Since the co-triad $e^I$ is orthonormal, it must satisfy the relation
\begin{equation}\label{13}
    g_{ab}=\eta_{IJ} e^I_a e^J_b=-l_a n_b- n_a l_b + m_a m_b,
\end{equation}
From the above relation and the properties of the null co-tetrad, we have the list constrains:
\begin{equation}\label{14}
    \eta_{IJ} l^I n^J=-1,\,\eta_{IJ}m^I m^J=1,\,others=0.
\end{equation}
There are nine parameters with six constrains, so it left only 3 free parameters which correspond to the SO(2,1) symmetry of the $e^I$.

We choose a gauge condition,
\begin{equation}\label{14a}
    l^0=l^1.
\end{equation}
Then we will get a solution for the constrain (\ref{14}):

\begin{equation} \label{15}
\left\{ \begin{aligned}
         l^0&=&l^1, l^2&=&0 \\
                  m^0&=&m^1, m^2&=&1\\
                  n^0&=&\frac{(m^0)^2+1}{2l^0},n^1&=&\frac{(m^0)^2-1}{2l^0},n^2&=&\frac{m^0}{l^0}.
\end{aligned} \right.
                          \end{equation}
Inserting the above solution to equation (\ref{5}), we will get the pull-back of the co-triad and the connection :

\begin{equation} \label{16}
\left\{ \begin{aligned}
         \underline{e^0_a}&=&\underline{e^1_a}=-l^0 n_a+m^0m_a,\,\underline{e^2_a}&=&m_a,\\
                  \underline{A^0_a}&=&\underline{A^1_a}=(\pi l^0-\epsilon m^0)n_a+(\alpha m^0-\mu l^0)m_a,\,\underline{A^2_a}&=&\alpha m_a -\epsilon n_a.
                          \end{aligned} \right.
                          \end{equation}
Remember that the index $I$ is rose and low by the matric $\eta_{IJ}=diag\{-1,+1,+1\}$, so $e_0=-e^0,e_1=e^1,e_2=e^2$.

 Putting the above relation to equation (\ref{3}), we will get
\begin{equation}\label{18}
\begin{split}
    \frac{1}{8\pi G}\int_{\Delta} \delta_2 \underline{e_I} \wedge \delta_1 \underline{A^I} =\frac{1}{8\pi G}\int_{\Delta} \delta_2 \underline{e^2} \wedge \delta_1 \underline{A^2} .\end{split}
\end{equation}
Now we consider a Lorentz boost $g=exp(\zeta)$ on the plane $(e^0,e^1)$. Under this boost, $e^2$ is unchanged, and the spin connection is changed into
\begin{equation}\label{19}
    A^2\rightarrow A^2+d\zeta.
\end{equation} This means that the connection $A^2$ is a SO(1,1) connection, and $e^2$ is in its adjoint representation. And this is what we need for a SO(1,1) BF theory.

Also notice that in order to keep the gauge condition (\ref{14a}), the internal symmetry has broken from SO(2,1) to the subgroup SO(1,1) which is the Lorentz group on the $\{e^0,e^1\}$ plane.

Since $\underline{e^2}=m_a$ is closed, we can locally define a 0-form $\tilde{B}$such that
\begin{equation}\label{19a}
    d\tilde{B}:=\underline{e^2}.
\end{equation}
 But the topology of the isolated horizon is non-trivial, that is, its first cohomology group is
\begin{equation}\label{20}
    H^1(\Delta)=H^1(S^1\times \mathbb{R})=H^0(S^1)=\mathbb{R}.
\end{equation}The integral of $m_a$ over any cross section yield $L_H:=2\pi R_{\Delta}$, where $L_H$ denote the horizon length. So $m_a$ is not in the zero class of $H^1(\Delta)$, the function must satisfy the constraint
\begin{equation}\label{20a}
    \oint_{S^1} d\tilde{B}=2\pi R_{\Delta}=L_H.
\end{equation}
On the other hand, $m_a,n_a$ are closed on the isolated horizon, and $\alpha,\epsilon$ are constant on the horizon\cite{ash2}, so $d A^2=0$, then the symplectic form of the boundary can be written as
\begin{equation}\label{18a}
    \Omega_S=\frac{1}{8\pi G}\int_{\Delta} \delta_{[2} \underline{e^2} \wedge \delta_{1]} \underline{A^2}=\frac{1}{8\pi G}\oint_{S^1}\delta_{[2} \tilde{B} \delta_{1]} A^2.
\end{equation}

So the full symplectic structure $\Omega$ on $\Gamma$ is given by :
\begin{equation}\label{8}
    \Omega|_{(A,e)}(\delta_1,\delta_2)=\frac{1}{8\pi G}\int_M (\delta_2 e_I \wedge \delta_1 A^I-\delta_1 e_I \wedge \delta_2 A^I) +\frac{1}{8\pi G}\oint_{S^1}(\delta_2 \tilde{B} \delta_1 A^2-\delta_1B' \delta_2 A^2)\equiv \Omega_V +\Omega_S.
\end{equation}
As we can see, the symplectic form split into the bulk term and the boundary term, so we can handle the quantization of the bulk and boundary degree of freedom separately. In the following section, we will show that the symplectic form on the boundary is precisely the symplectic form of a topological BF theory on the isolated horizon.
\section{2D BF theory}
The action of a 2D topological BF theory\cite{bf1,bf2} on a manifold $\Delta$ is given by
\begin{equation}\label{9}
    S[B,A]=\int_{\Delta} Tr [B F]=\int_{\Delta} B^I F_I,
\end{equation}
where $A$ is a $G-$valued connection field, $F$ its field strength 2-form, and $B$ a scalar field in the adjoint representation of $G$. In our case, $G$ is the  abelian group SO(1,1), so we will omit $I$.

The field equation are
\begin{equation}\label{10}
    \tilde{F}:=dB=0,\quad F=dA=0.
\end{equation}
Assume the manifold has the topology $S^1 \times \mathbb{R}$. From the action (\ref{9}) we can get the symplectic form of BF theory\cite{mm1}:
\begin{equation}\label{11}
    \Omega|_{(B,A)}(\delta_1, \delta_2)=\oint_{S^1}(\delta_2 B \delta_1 A-\delta_1 B  \delta_2 A).
\end{equation}
Comparing equation (\ref{11}) with the boundary term of equation of (\ref{8}), we can see that, after making the identity
\begin{equation}\label{12}
    B \leftrightarrow \frac{1}{8\pi G}\tilde{B},\, A \leftrightarrow A^2,
\end{equation}
they have the same form, so we get the result that the degrees of freedom at the horizon can be described by an SO(1,1) BF theory.

From the experience of black hole in 3+1 dimension, we know that the relevant field theory should be topological field theory with punctures, so we have to add some particles, and the full theory will be\cite{enpp1}:
\begin{equation}\label{21}
    S_f=S(B,A)+S_p(A,\chi_1,\cdots,\chi_n)=\int_{\Delta} B dA+\sum_{p=1}^n \lambda_p \int_{c_p} (\chi_p^{-1}d \chi_p+A).
\end{equation}
where $c_p\subset \Delta$ are the particle world line, $\lambda_p$ coupling constant, and $\chi_p$ are SO(1,1) valued d.o.f of the particle. The gauge symmetry for the full action are
\begin{equation}\label{22}
    A\rightarrow A - d\alpha, B\rightarrow B, \chi_p\rightarrow e^{\alpha(x)}\chi_p,
\end{equation}where $e^{\alpha(x)}\in SO(1,1)$. As a Lie group, $SO(1,1)\cong \mathbb{R}$.

The equation of motion will be given by:
\begin{equation}\label{23}
    F=dA=0,\quad \tilde{F}=dB=\sum_{p=1}^n\lambda_p \delta(x,x_p).
\end{equation}
This is to say, after adding the punctures, the connection is still flat everywhere, and the $B$ field curved due to the punctures.

Following paper \cite{enpp1} we quantize this punctured BF theory. In order to perform the canonical analysis we assume that $\chi_p(r)=exp(r_p)$. Under the left action of the group we have
\begin{equation}\label{29c}
    exp(\alpha) \chi_p=exp((\alpha+r_p))
\end{equation}the infinitesimal version of the previous action is
\begin{equation}\label{30c}
     \chi_p=\frac{\partial \chi_p }{\partial r}.
\end{equation}

Define the momentum $S_p$ as the conjugate momentum of $r$ then it satisfy the Poisson bracket:
\begin{equation}\label{31c}
    \{r_p, S_p\}= 1.
\end{equation}
Explicit computation shows that $S_p= \lambda_p$. So we have a primary constrain per particle
\begin{equation}\label{32c}
    \Psi(S_p, \chi_p)\equiv S_p- \lambda_p = 0.
\end{equation}

In summary, the phase space of each particle is $T^*(SO(1,1))\cong T^* \mathbb{R}$, where the momentum conjugate to $\chi_p$ is given by $S_p$ satisfying the Poisson bracket (\ref{31c}).

Following \cite{enpp1}, we perform the canonical analysis. The event horizon has the form $\Delta=H\times \mathbb{R}$, where $H$ is a one-dimensional manifold with circle topology. Let $x^a=(t,\phi)$ on the $\Delta$, where $t$ is the non-compact coordinate along $\mathbb{R}$ and $\phi \in [0,2\pi]$. Then the action can be written as:

\begin{equation}\label{32b}\begin{split}
    S=\int_R dt \int_H d\phi [ B \partial_t A_{\phi} + A_t (\tilde{F}-\sum_{p=1}^n \delta(x,x_p)\lambda_p)+\sum_{p=1}^n\delta(x,x_p)\lambda_p \partial_t r_p].
\end{split}\end{equation}

The kinematical terms of (\ref{32b}) involving time derivative determine the Poisson bracket:
\begin{equation}\label{33b}\begin{split}
    \{A(\phi),B(\phi')\}=\delta(\phi,\phi'),\\
    \{r_p,\lambda_{p}\}=1.
\end{split}\end{equation}
The Lagrange multipliers $A_t$ enforce the following constrains:
\begin{equation}\label{33a}
 \tilde{F}=\sum_{p=1}^n \delta(x,x_p)\lambda_p,
\end{equation}
together with the constraint (\ref{32c})
\begin{equation}\label{34a}
    S_p- \lambda_p=0.
\end{equation}
Since the theory is topological, the non trivial d.o.f are only present at punctures. Note that since $i S_p$ generated the SO(1,1) symmetry, $\lambda_p$ are quantized according to $\lambda_p=a_p$ where $a_p$ is a real number labeling the unitary irreducible representation of SO(1,1).

From now on we denote the $\mathcal{H}^{BF}(a_1,\cdots,a_n)$ as the Hilbert space of the BF theory associated with a fixing choice of the momentum number $a_p$ at each punctures $p\in \gamma \cap H$ and a state by $|\{a_p\}_1^n;\cdots>$.
\section{The Bulk Theory}
In this section, we summarize the properties of the LQG in 2+1 dimension\cite{rov2}.

In 2+1 dimension, the Lorentzian general relativity can be formula as follows. The gravitational field is represented by an SO(2,1) connection $A_{\mu}^I(x)$ and a triad $e_{\mu}^I$. Here $\mu=0,1,2$ is a space-time index, and $I=0,1,2$ is an internal index, labeling a basis in the Lie algebra $so(2,1)$. We will work in a space-time of signature $(-,+,+)$, so we raise and lower the internal indices using the flat metric $\eta_{IJ}=diag(-,+,+)$. The action is given by
\begin{equation}\label{35}
    S[e,A]=\frac{1}{8\pi G}\int (e^I \wedge F_I-\frac{\Lambda}{6}\epsilon^{IJK}e_I\wedge e_J \wedge e_K).
\end{equation}
We can perform the Hamiltonian analysis, by choose $x^0$ as the time evolution parameter and $x^a=(x^1,x^2)$ as coordinates of the initial surface $\Sigma$. Then the canonical variables are $A_a^I(x)$, and their conjugate momentum $\pi^a_I=\frac{1}{8\pi G}\eta_{IJ}\epsilon^{ab}e^J_b(x)$. The fundamental Poisson bracket is therefore
\begin{equation}\label{36}
    \{A^I_a(x), e^J_b(y)\}=8\pi G \epsilon_{ab} \eta^{IJ} \delta^{(2)} (x,y).
\end{equation}

Now we can apply the loop quantization procedure. For detail see paper\cite{rov2}, and we just summarize some results. The kinematical Hilbert space is given by the SO(2,1) spin network functional.

For SO(2,1) group\cite{rov2}, we denote $X_I$ be the generators of a linear representation of the group. They are linear operators that satisfy:
\begin{equation}\label{17}
  [X_0,X_1]=-X_2,\,[X_1,X_2]=X_0,\,[X_2,X_0]=-X_1.
\end{equation}
One can check that they are the right sign for the symmetry group SO(2,1) of a $(-,+,+)$ Lorentz space. For unitary representation, the operator $iX_I$ are hermitian. The Casimir operator is given by
\begin{equation}\label{17a}
 Q=(X_0)^2-(X_1)^2-(X_2)^2.
\end{equation}

An important result is the spectrum of the length operator. In 3+1 dimension it is well known that the spectrum of the area operator is discrete, and in 2+1 dimension the length take the same role as area in 3+1 dimension. But in surprise, the result is different from expectation. The spectrum of the time-like intervals is discrete, but for the space-like interval continuous. The length operator has the following form:
\begin{equation}\label{38}
    \hat{L}_c \Psi^{(\mathcal{I})}=8\pi G \sqrt{q^{(\mathcal{I})}} \Psi^{(\mathcal{I})},
\end{equation}
where $q^{(\mathcal{I})}$ is the value of the Casimir operator for SO(2,1) in the representation $\mathcal{I}$.

In this paper, we take the second length spectrum appeared in the paper\cite{rov2} because it fits better with other approaches. And also we use the double cover group SU(1,1) instead of the SO(2,1). Then the length spectrum become
\begin{equation}\label{39}
\left\{ \begin{aligned}
         L_q&=&8\pi G q ,\,for \, space-like, \, \mathcal{C}_{q>0}\\
                  T_{\epsilon,n}&=&8\pi G \epsilon (n-1/2),\ for \, \mathcal{D}_{n\geq1}^{\epsilon=\pm},\, n\in \mathbb{N}/2.
                          \end{aligned} \right.
\end{equation}
That is, the spectrum of the time-like interval is half-integer multiplication of the fundamental length and the space-like continuous. A spin network of SU(1,1) is character by two numbers $|q, m>$, where $q$ is the eigenvalue of the Casimir operator and $m$ the eigenvalue of the hermitian operator $-iX_0$.

For space-like interval $H=\Delta \cap M$, the spectrum of the length operator is continuous. From the experience of 4D black hole, we know that the area constrain play an essential role in the state counting of the black hole. It reduce the dimension from the infinite to finite because of the quantization of the area spectrum. And in 3D case, since the spectrum is continuous, it play almost no role, and the dimension is infinite and is not the result we want.

Another important result will be the action of the triad operator $e^I$. It is easy to show that\cite{rov2}
\begin{equation}\label{41}
    \hat{e}^I(x)=-i8\pi G \sum_{p\in \gamma\cap H}\delta(x,x_p) \hat{X}^I(p).
\end{equation}
where $X^I$ are the generators of the SU(1,1) group. The only left component in the boundary symplectic form is $e^2$. When it act on the SO(2,1) spin network $|q,m>$, it is not diagonal. So we must use another basis of the spin network which can diagonal $X_2$. The basis is given\cite{mr1,dav} by $|q,l>$ where $l \in \mathbb{R}$. In the following, we will denote the spin network bases which diagonalized both the Casimir operator and the third component as $|q,l>$ which $q,l \in \mathbb{R}$.

In order to calculate the entropy of the BTZ black hole, we also need the boundary condition on the horizon $H=\Delta \cap M$, which topologically is $S^1$. Since it is 1 dimension manifold, a natural choices for the boundary condition would be a equation for 1-form. According to the equation (\ref{12}) and (\ref{19a}), we propose the following isolated horizon boundary condition:
\begin{equation}\label{24}
    \tilde{F}\circeq\frac{1}{8\pi G}e^2.
\end{equation}
where $\circeq$ means equal on the $H=\Delta \cap M$. Or more precisely, its integral form
\begin{equation}\label{24b}
    \oint_S \tilde{F}\circeq\frac{1}{8\pi G} \oint_S e^2,
\end{equation}
since only the integral form have well-defined meaning in quantum theory. We will show that this boundary condition can give the desired result.

The following step are as same as in the 3+1 dimension\cite{enpp1}. The full Hilbert space will be the tensor product of the bulk and boundary Hilbert space. And the quantum version of the boundary condition will be
\begin{equation}\label{24a}
    (Id\otimes \oint_S \hat{\tilde{F}}-\frac{1}{8\pi G}\oint _S \hat{e}^2\otimes Id)(\Psi_v \otimes \Psi_b)=0.
\end{equation}

 One consider the bulk Hilbert space $\mathcal{H}_{\gamma}$ defined on a graph $\gamma \subset M$ with end points on $H$, denoted $p\in \gamma \cap H$. The quantum operator associated with $e^2$ in (\ref{24}) is
\begin{equation}\label{25}
\begin{split}    \oint_S \hat{e}^2(x)|\{q_p,l_p\}_1^n;\cdots>=-i8\pi l_{Pl} \sum_{p\in \gamma\cap H} \hat{X}^2(p)|\{q_p,l_p\}_1^n;\cdots >\\=8\pi l_{Pl}\sum_{p\in \gamma\cap H}l_p|\{q_p, l_p\}_1^n;\cdots>,\end{split}
\end{equation}
  Inserting this equation to the boundary condition (\ref{24}) we can get
 \begin{equation}\label{26}
    \oint_S \hat{\tilde{F}}|\{a_p\}_1^n;\cdots>= \sum_{p\in \gamma\cap H}l_p|\{a_p\}_1^n;\cdots>,
 \end{equation}
 which is exactly the integral form of the equation (\ref{33a}) for the $B$ field in the presence of particles. It also give a relation between charge number in the SO(1,1) BF theory and the magnetic number in the SO(2,1) bulk theory: $a_p= l_p$. It shows that we choose right isolated horizon boundary condition.

 The horizon length operator will have the following eigenvalues:
 \begin{equation}\label{27}
    \hat{L}_H |\{j_p,m_p\}_1^n;\cdots>=8\pi l_{Pl}  \sum_{p=1}^n q_p|\{q_p,l_p\}_1^n;\cdots >.
 \end{equation}
 As in 3+1 dimension, the entropy is given by $S=log(\mathcal{N})$ where $\mathcal{N}$ is the number of horizon states $|{a_p}_1^n;\cdots>$. There are some constraints on those states. One is familiar to us: the length of the horizon is $L_H$, that is
 \begin{equation}\label{44}
   8\pi l_{Pl}  \sum_{p=1}^n q_p= L_H.
 \end{equation}

 And there is another global constraint from the boundary condition (\ref{20a}):
 \begin{equation}\label{45}
   8\pi G \sum \lambda_p=8\pi G \sum a_p= 8\pi G \sum l_p=8\pi l_{Pl} \sum l_p=L_H.
 \end{equation}

On the kinematical Hilbert space spanned by the SO(2,1) spin network states $|q,l>$, the $q,l \in \mathbb{R}$, so there are infinite number of states that satisfying the above two constraints. Unlike the SU(2) case, which there is a relation $-j\leq m \leq j$ between the two quantum numbers, there is no such a relation for $q$ and $l$.

But remember that this is just on the kinematical level. And the physical Hilbert space is what we need. In three dimension\cite{car2}, there is no local degree of freedom for gravity, and just few global degrees of freedom. This is very different from the gravity in four dimension which does have local d.o.fs. On the kinematical level, the difference between 3D and 4D is small, but on the physical level, they have huge difference. And those difference can be due to the Hamiltonian constraint. Since we don't have well understanding of the Hamiltonian constraint and the physical Hilbert space for 3D quantum gravity, we make some assumptions on the physical states which make us to move on.

 We got our idea from the paper\cite{bn1} which consider the higher derivative Lovelock gravity theory. The key observation of this paper is that "in a loop quantization of a generalized gravity, the analog of the area operator turns out to be measure, morally speaking, the Wald entropy rather than the area." Actually it is the flux operator that appeared in the Wald entropy formula, not the area operator. So we think that the flux operator play a more fundamental role than the area operator. Following this idea, we make the following 'quantized flux' assumption:
for physical states, the magnetic number $l$ are quantized according to
\begin{equation}\label{47}
    l_m=\alpha m,\, m \in N^+.
\end{equation}
where $\alpha$ is a constant.

So we only have one constraint (\ref{45}). Denote $a=\frac{L_H}{8\pi l_{Pl}\alpha}$. Then the problem reduced to a Combinatorics problem: partition of a integer into ordered positive integer. The result is well known:  the number of the states is given by
\begin{equation}\label{48}
    \mathcal{N}=2^{a-1},
\end{equation}
Then the entropy for the BTZ black hole is
\begin{equation}\label{49}
    S=log(\mathcal{N})\approx \frac{log(2)L_H}{8\pi \alpha l_{Pl}}-log(2).
\end{equation}

If we set $\alpha=log(2)/(2\pi)$, we will get the $1/4$ coefficient of the Bekenstein-Hawking area law. From this point of view, the $\alpha$ is like the Barbero-Immirzi parameter in loop quantum gravity in 4D. But we don't have any logarithmic correction which is a problem.

After choosing $\alpha=log(2)/(2\pi)$, the spectrum of the length operator for the horizon is given by
\begin{equation}\label{52}
    L_n=8\pi l_{Pl} \alpha n=4 log(2) n l_{Pl}.
\end{equation}

We must stress that this is just a simple solution to the constraints. There are maybe other solutions that seems very different from ours. But we learn a lesson that to explain the entropy of the BTZ black hole, we must going into the dynamics of the quantum gravity, not only on the kinematical level, which is the case in LQG in 4D. The black hole physics give some clues on the dynamics of the quantum gravity, just people expect.
 \section{Conclusion}
 In this paper, we apply the method developed in loop quantum gravity to BTZ black hole. We analysis its symplectic structure and find that the symplectic form split into two terms: the bulk term and the boundary term, so we can quantized the bulk and boundary part separately. We also find that after a gauge fixing the boundary  theory is a 2D SO(1,1) BF theory.

 To calculate the entropy of the black hole, we make some assumption on the nature of the physical states. We give an estimate of the entropy of the BTZ black hole, and find that it obey the Bekenstein-Hawking area law with the famous 1/4 factor. But we don't get the right sub-leading term.

 A lesson we learned is that the black hole physics will give some hints on the dynamics of the quantum gravity.

 Chern-Simons theory can only be defined on odd dimension space-time, which limit its application to black hole physics.
  On the other hand, the BF theory can be defined on any dimension, so it maybe a better choice. Applying the BF theory to balck hole in 4 dimension
   and higher dimensional $(D\geq 5)$ black hole is under investigation.
 \acknowledgments
 The author would like to thank the loop quantum gravity team in Beijing Normal University. This work is supported by the NSFC (Grant No. 11235003) and
  the Research Fund for the Doctoral Program of Higher Education of China. \bibliography{btz1}
\end{document}